%% file: main.tex
\documentclass{INTERSPEECH2023}
\usepackage{amsmath,epsfig}

\let\OLDthebibliography\thebibliography
\renewcommand\thebibliography[1]{
  \OLDthebibliography{#1}
  \setlength{\parskip}{0pt}
  \setlength{\itemsep}{0pt plus 0.3ex}
}
\usepackage{multirow}
\usepackage{booktabs}
\usepackage{cite}

\usepackage{bm}
\usepackage{graphicx}
\usepackage{subfigure}
\ninept

\title{AFL-Net: Integrating Audio, Facial, and Lip Modalities with a Two-step Cross-attention for Robust Speaker Diarization in the Wild}
\name{Yongkang Yin$^{1,\dagger}$\thanks{$^{\dagger}$This work was done when Yongkang Yin was an intern at ARC Lab, Tencent PCG.}, Xu Li$^{2,*}$, Ying Shan$^2$, Yuexian Zou$^{1,*}$\thanks{* Corresponding authors.}}
\address{
    $^1$ ADSPLAB, School of ECE, Peking University, China\\
    $^2$ ARC Lab, Tencent PCG\\
    \small{
        yinyongkang@stu.pku.edu.cn, \{nelsonxli, yingsshan\}@tencent.com, zouyx@pku.edu.cn
    }
}

\begin{document}

\maketitle
\input{speaker-diarization/0-abstract}
\input{speaker-diarization/1-introduction}
\input{speaker-diarization/2-methodology}

\input{speaker-diarization/3-experiments}
\input{speaker-diarization/4-conclusion}

\bibliographystyle{IEEEtran}
\bibliography{main}

\end{document}

%% file: speaker-diarization/0-abstract.tex
\begin{abstract}

Speaker diarization in real-world videos presents significant challenges due to varying acoustic conditions, diverse scenes, the presence of off-screen speakers, etc. This paper builds upon a previous study (AVR-Net) and introduces a novel multi-modal speaker diarization system, AFL-Net. The proposed AFL-Net incorporates dynamic lip movement as an additional modality to enhance the identity distinction. Besides, unlike AVR-Net which extracts high-level representations from each modality independently, AFL-Net employs a two-step cross-attention mechanism to sufficiently fuse different modalities, resulting in more comprehensive information to enhance the performance. Moreover, we also incorporated a masking strategy during training, where the face and lip modalities are randomly obscured. This strategy enhances the impact of the audio modality on the system outputs. Experimental results demonstrate that AFL-Net outperforms state-of-the-art baselines, such as the AVR-Net and DyViSE.


\end{abstract}

%% file: speaker-diarization/1-introduction.tex
\section{Introduction}
\label{sec:intro}

\begin{figure*}[t]
	\includegraphics[width=\textwidth, height = 8.5cm]{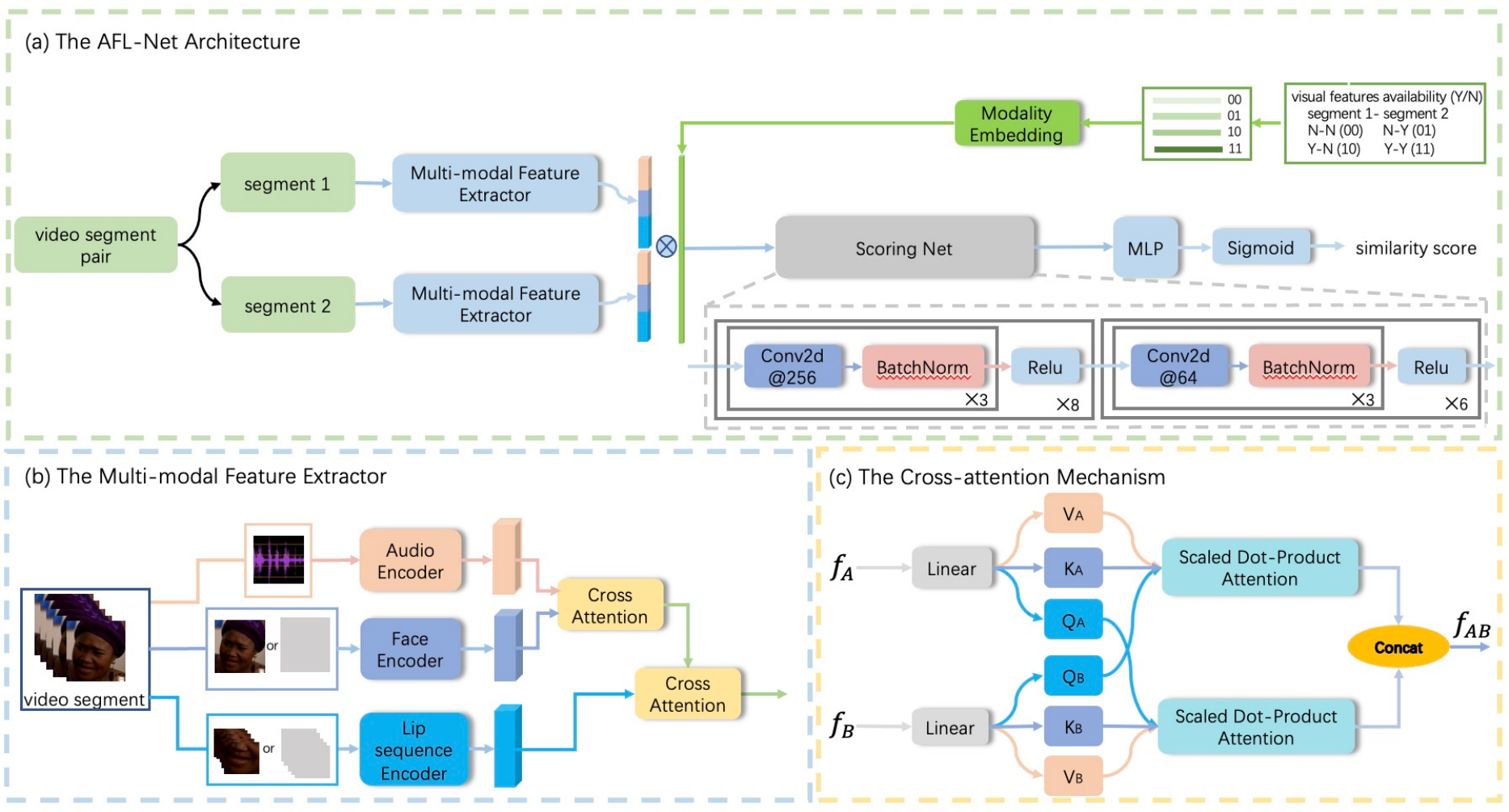}
 	\caption{The proposed system architecture. (a) represents the AFL-Net architecture, where the MLP indicates several linear layers. (b) illustrates the multi-modal feature extractor and (c) demonstrates the cross-attention mechanism.}
	\label{fig: architecture}
\end{figure*}

Speaker diarization is a cutting-edge technology that identifies ``who spoke when'' in audio or video recordings~\cite{park2022review}. This task involves segmenting the input into homogeneous regions based on speaker identity, playing a crucial role in applications like transcription services, meeting analysis and broadcast news indexing.

Traditional speaker diarization primarily utilized audio features like spectral characteristics and prosodic patterns for speaker identification and segmentation. Early works~\cite{landini2022bayesian} followed a multi-stage approach, encompassing speech enhancement, voice activity detection (VAD), speaker feature extraction, similarity scoring, and clustering.
Subsequent studies~\cite{fujita2019end, zhang2019fully} employed deep learning networks for end-to-end speaker diarization, with later research addressing associated issues, such as uncertain speaker count~\cite{horiguchi2022encoder}.
In recent years, multi-modal speaker diarization has integrated audio and visual features to improve performance. These studies employed various methods, such as CNN enhancement~\cite{fanaras2022audio}, self-supervised learning~\cite{ding2020self}, etc.
Some works incorporated additional features like speaker i-vectors~\cite{he2022end} and localization features~\cite{chung2019said,wuerkaixi2022dyvise} to enhance system performance.
Other studies focused on specific scenarios, such as real-world conferences and in-the-wild settings~\cite{wang2022cross,zhongcong2021ava}. 
Despite achieving reasonable performance, these methods encountered limitations in complex and noisy environments, resulting in suboptimal outcomes.

In real-world scenarios like movies, challenging scenes with off-screen speakers hinder the direct use of audio-visual speaker diarization (AVSD) systems. The AVR-Net~\cite{zhongcong2021ava} was proposed by using a modality embedding that reflects face visibility, making AVSD adaptable to situations where the face is not visible.
While the AVR-Net demonstrated satisfactory performance in real-world scenarios, this study suggests that there is room for further enhancements to improve its overall performance.
1) Besides the face and audio modalities that are considered by the AVR-Net to compute the speaker similarity between segments, lip movement is another key modality but not taken into consideration by the AVR-Net. The dynamic lip movement could reflect a person's pronunciation styles, which are complementary to face and audio modalities and can enhance the discrimination of identities~\cite{he2022end}.
2) In the AVR-Net, different modalities may not be sufficiently fused according to the network architecture. The face and audio modalities are not orthogonal and should be conditioned on each other to generate high-level representations, rather than generating the representations independently and simply concatenating these representations of each modality for segment scoring as in \cite{zhongcong2021ava}.

To address the issues above, this work proposes a novel multi-modal speaker diarization network, named AFL-Net, that extends the AVR-Net in the following aspects: 1) Besides the face and audio modalities, we leverage the dynamic lip movement as an additional modality to improve the discrimination of identities; 2) We design a two-step cross-attention fusion module to generate high-level representations of each modality while conditioning on other modalities; 3) Since the visual modalities have the potential to be partially absent (e.g. partially obscured face or incomplete lip movement on side face), we propose a masking strategy to randomly mask out the face and lip movement modalities during training, which increases the impact of the audio modality on system outputs.

A concurrent work, DyViSE \cite{wuerkaixi2022dyvise}, bears similarity to ours. However, the proposed AFL-Net distinguishes itself in several ways: 1) The AFL-Net employs a two-step cross-attention method to fuse three modalities, whereas \cite{wuerkaixi2022dyvise} only applies attention between the lip and audio modalities; 2) \cite{wuerkaixi2022dyvise} depends on three pre-trained models for embedding computation and uses cosine similarity for segment scoring. In contrast, our method directly trains a scoring network to output similarity scores between segments; 3) The AFL-Net adopts an additional masking strategy during training. The dedicated architecture of AFL-Net enables it to outperform DyViSE, as will be demonstrated in Section~\ref{subsec: expt_1}.

The contributions of this work are as follows: 1) Leveraging the dynamic lip movement as an additional modality to enhance the system's identity discrimination; 2) Introducing a two-step cross-attention module to sufficiently fuse different modalities for similarity scoring; 3) Proposing a masking strategy to randomly mask out the visual modalities during training, which leads the system to make decisions more on the audio information and experimentally achieve better speaker diarization performance.

The rest of this paper is organized as follows: Section~\ref{sec:Method} illustrates the proposed system. Experimental setup and experiment results are demonstrated in Section~\ref{sec:expt-setup} and \ref{sec:expt-rst}, respectively. We conclude this work in Section~\ref{sec:conclusion}.

%% file: speaker-diarization/2-methodology.tex
\section{The Proposed Method}
\label{sec:Method}

The proposed AFL-Net adopts the general architecture of the AVR-Net \cite{zhongcong2021ava}, which trains the network to assign an identity similarity score to each pair of two video segments. During inference, a testing video is initially divided into shorter segments, and the network predicts an identity similarity score between each pair of segments. 
Subsequently, a clustering algorithm, such as agglomerative hierarchical clustering (AHC), is applied to group segments with high scores, indicating that they likely belong to the same person.
This work modifies the scoring network of \cite{zhongcong2021ava} while adopting the same AHC algorithm for clustering.

The proposed AFL-Net is demonstrated in Fig.~\ref{fig: architecture}.
In Fig.~\ref{fig: architecture}(a), given a pair of two video segments, a multi-modal feature extractor is applied to extract high-level representations for each segment independently, then the two representations are concatenated together for scoring. 
Since there are instances where visual features are absent, we utilize 4 learnable modality embeddings \cite{zhongcong2021ava} to indicate whether the visual features are available or not for each segment. These embeddings are randomly initialized and optimized during training.
The concatenated representations are multiplied with the selected modality embedding, then fed into the scoring network to generate the similarity score for this pair of segments.

\subsection{Multi-modal feature extractor}

The multi-modal feature extractor is illustrated in Fig.~\ref{fig: architecture}(b). Besides the face and audio modalities adopted in \cite{zhongcong2021ava}, this work additionally leverages the dynamic lip movement as the third modality to be fused in the model because lip movement could reflect a person's pronunciation styles, which are complementary to face and audio modalities and can enhance the discrimination of identities~\cite{he2022end}.
We utilize three open-sourced pre-trained encoders to extract audio\cite{chung2020defence}, face\cite{deng2019arcface}, and lip movement\cite{martinez2020lipreading} features, respectively.
All of them have a ResNet-like\cite{he2016deep} network architecture and have been trained on large-scale datasets, yielding impressive recognition performance.

Furthermore, to effectively integrate information from different modalities, we employ a two-step cross-attention mechanism. This mechanism allows for the generation of high-level representations for each modality while considering queries from other modalities.
The structure of a single cross-attention is shown in Fig.~\ref{fig: architecture}(c). Suppose that we have two modality features $f_{A}$ and $f_{B}$, the key, query and value vectors of each modality are first extracted through a linear layer, denoted by $K_{A}$, $Q_{A}$, $V_{A}$, $K_{B}$, $Q_{B}$ and $V_{B}$, respectively. The cross-attended fusion of these two modalities is derived by Eq.~\ref{eq: two_modality_cross_attention}:
\begin{align}
    f_{AB} = \mathrm{softmax} (\frac{Q_{B} K_{A}^{T}}{\sqrt{d_A}}) V_{A} \oplus \mathrm{softmax} (\frac{Q_{A} K_{B}^{T}}{\sqrt{d_B}}) V_{B} \label{eq: two_modality_cross_attention}
\end{align}
where $f_{AB}$ represents the fused feature of the modality $A$ and $B$, $d_A$ and $d_B$ are the dimensions of the corresponding key vectors, and $\oplus$ denotes the concatenation operation.

Please note that there are multiple fusion implementations for three modalities, including a one-step three-modality fusion or a two-step two-modality fusion.
In this study, four fusion implementations were explored, and experimental results indicate that the optimal implementation involves initially fusing the audio and face modalities, followed by the inclusion of the lip movement modality in the second step, as illustrated in Fig.~\ref{fig: architecture}(b).
Further details will be discussed in Section~\ref{subsec: expt_3}.
The implementation is illustrated by Eq.~\ref{eq: three_modality_cross_attention_1} and \ref{eq: three_modality_cross_attention_2}:
\begin{align}
    f_{AF} &= \mathrm{softmax} (\frac{Q_{F} K_{A}^{T}}{\sqrt{d_A}}) V_{A} \oplus \mathrm{softmax} (\frac{Q_{A} K_{F}^{T}}{\sqrt{d_F}}) V_{F} \label{eq: three_modality_cross_attention_1} 
\end{align}
\begin{align}
    f_{AFL} &= \mathrm{softmax} (\frac{Q_{L} K_{AF}^{T}}{\sqrt{d_{AF}}}) V_{AF} \oplus \mathrm{softmax} (\frac{Q_{AF} K_{L}^{T}}{\sqrt{d_L}}) V_{L} \label{eq: three_modality_cross_attention_2}
\end{align}
Finally, $f_{AFL}$ is the fused feature of the three modalities.

\subsection{Masking strategy}

Given that visual information in videos, such as face and lip movement modalities, can be partially absent (e.g., due to obscured faces or incomplete lip movements on the side face), systems that heavily rely on visual inputs may encounter a significant performance decline in such challenging scenarios. To mitigate this issue, we propose a masking strategy, where the face and lip movement modalities are randomly masked out during training. This strategy aims to amplify the impact of the audio modality on the system outputs, compensating for the potential absence of visual cues.
In our approach, we randomly mask out the visible face and lip movement modalities at a rate of 0.3 during training. This process also includes updating the corresponding modality embedding. It is important to note that this masking strategy is only applied during training, and the testing data remains unaltered for inference.


\subsection{Agglomerative hierarchical clustering}
Following \cite{zhongcong2021ava}, we utilize the agglomerative hierarchical clustering (AHC) algorithm \cite{day1984efficient} for speaker clustering.
AHC is a bottom-up clustering method where each data point begins as an individual cluster. It iteratively merges the closest clusters until the largest similarity score reaches a predefined threshold.
In our approach, AFL-Net predicts identity similarity scores between segments within a video. Using these scores, the AHC method groups segments into clusters, each representing a specific person's identity. Notably, this method eliminates the need for pre-specifying the number of speakers in a video.


%% file: speaker-diarization/3-experiments.tex
\section{Experimental Setup}
\label{sec:expt-setup}
\subsection{Datasets}

The experiments involve three datasets: AVA-Audio-Visual Diarization (AVA-AVD)~\cite{zhongcong2021ava}, VoxCeleb1~\cite{nagrani2017voxceleb} and VoxCeleb2~\cite{chung2018voxceleb2}.
The AVA-AVD~\cite{zhongcong2021ava} dataset comprises 351 video clips from 117 movies, each lasting 5 minutes with a maximum of 24 speakers.
VoxCeleb1 and VoxCeleb2 are two popular datasets for speaker classification tasks, totally containing 1M+ video segments from over 7,000 celebs.
The AVA-AVD dataset is much more challenging due to the diverse scenes and complicated acoustic conditions~\cite{zhongcong2021ava} but relatively small, this work utilizes it in both training and evaluation. 
To further confirm the effectiveness of the proposed model trained on a larger dataset, the VoxCeleb1 and VoxCeleb2 are utilized as the extra training data because of their much larger scales.

\subsection{Model and training configurations}
Following \cite{zhongcong2021ava}, we partition the 16-bit 16kHz mono-channel active speaker audio segments, obtained after a speech enhancement process and voice activity detection (VAD)~\cite{sun2018speaker, sredojev2015webrtc}, into segments of 0.5 seconds each.
Concurrently, we extract facial images and lip sequence images within the same audio time frame, leveraging existing open-source tools~\cite{deng2020retinaface, bulat2017far, haliassos2021lips}.
From each time slot, we randomly select one face image and ten lip images. In instances where visual features are absent, a zeroing operation is performed as a placeholder, as shown in Fig.~\ref{fig: architecture}(b).

The loss function for training the AFL-Net is defined as:
\begin{align}
    \mathcal{L}_{\mathrm{MSE}}=\sum_{i =1}^{N}\sum_{j=1}^{N} (s_{i,j}-1(y_{i}
==y_{j}))^2 \label{eq: loss_func}
\end{align}
where N represents the batch size, $y_{i}$ and $y_{j}$ represent the identity labels, and $s_{i,j}$ is the predicted score by the AFL-Net. The AFL-Net is trained to predict 1 if two segments are from the same person and 0 vice versa.
The Adam optimizer is utilized to update model parameters, with an initial learning rate of $5\times 10^{-4}$.
After 6k iterations of training, the model is evaluated on the validation set of the AVA-AVD dataset every 500 iterations. The model parameters that achieve the best performance on the validation set are selected for inference. 
For the clustering backend, we search the threshold within the range from 0.1 to 0.3 with a step size of 0.01. The threshold that yields the minimum DER in validation is recorded for the testing period.

\subsection{Evaluation metric}
There are four commonly used evaluation metrics in speaker diarization~\cite{anguera2012speaker}: missing rate (MR), false alarm rate (FAR), speaker error rate (SpkErr), and diarization error rate (DER). DER is a comprehensive measure that combines MR, FAR, and SpkErr, calculated using Eq.~\ref{eq: der}.
\begin{align}
    DER = \frac{T_{MS} + T_{FA} + T_{SPKE}}{T_{total}} \label{eq: der}
\end{align}
where $T_{total}$ represents all the testing frames, $T_{MS}$, $T_{FA}$ and $T_{SPKE}$ represent the misclassified frames that cause the miss, the false alarm and the speaker error, respectively.

The AVR-Net \cite{zhongcong2021ava} and the DyViSE \cite{wuerkaixi2022dyvise} are selected as baseline models for performance comparison. Both the AVR-Net and the AFL-Net utilize the same VAD tool, resulting in identical performance in terms of MR (2.55\%) and FAR (0). Hence, we only demonstrate the SpkErr and the DER results for comparing the AFL-Net and the AVR-Net in Table~\ref{tab:main}. 
The DyViSE \cite{wuerkaixi2022dyvise} utilizes a stronger VAD tool \cite{tian2022royalflush}, resulting in a slightly lower MR of 1.98\%.
Furthermore, the audio encoder in DyViSE utilizes a substantial 310M-parameter WavLM model, whereas our approach only employs a modest 9M parameters. To ensure a fair comparison, we also integrate the WavLM model into our audio encoder, which leads to improved performance. The comparison of SpkErr and DER results between AFL-Net and DyViSE can be found in Table~\ref{tab:main2}.

\section{Experiment Results}
\label{sec:expt-rst}


\begin{table}[t]
\centering
\caption{Performance comparison between AVR-Net and AFL-Net. AVA-AVD refers to models trained solely on the AVA-AVD dataset, while w/ Extra Data indicates models trained on a combination of Voxceleb1, Voxceleb2, and AVA-AVD.}
\begin{tabular}{ccccc}
\toprule[1pt]
\multirow{2}{*}{\textbf{Models}} & \multicolumn{2}{c}{\textbf{AVA-AVD}} & \multicolumn{2}{c}{\textbf{w/ Extra Data}} \\ 
& \multicolumn{1}{c}{\textbf{SpkErr} $\downarrow$} 
& \multicolumn{1}{c}{\textbf{DER} $\downarrow$} 
& \multicolumn{1}{c}{\textbf{SpkErr} $\downarrow$} 
& \multicolumn{1}{c}{\textbf{DER} $\downarrow$}\\ \midrule[1pt]

AVR-Net & 24.88\% & 27.43\% & 18.58\% & 21.13\%\\
AFL-Net & \textbf{21.10\%} & \textbf{23.65\%} & \textbf{17.10\%} & \textbf{19.65\%}\\
\bottomrule[1pt]
\end{tabular}
\vspace{-1em}
\label{tab:main}
\end{table}

\begin{table}[t]
\centering
\caption{Performance comparison between DyViSE and AFL-Net (with WavLM) on AVA-AVD dataset.}
\setlength{\tabcolsep}{4.5pt}
\begin{tabular}{ccccc}
\toprule[1pt]
\multirow{1}{*}{\textbf{Models} }
& \multicolumn{1}{c}{\textbf{FAR} $\downarrow$} 
& \multicolumn{1}{c}{\textbf{MR} $\downarrow$} 
& \multicolumn{1}{c}{\textbf{SpkErr} $\downarrow$} 
& \multicolumn{1}{c}{\textbf{DER} $\downarrow$}\\ \midrule[1pt]
DyViSE \cite{wuerkaixi2022dyvise} & 0.0 & \textbf{1.98\%} & 20.86\% & 23.46\%\\
AFL-Net + WavLM & \textbf{0.0} & 2.55\% & \textbf{19.57\%} & \textbf{22.12\%}\\
\bottomrule[1pt]
\end{tabular}
\vspace{-1em}
\label{tab:main2}
\end{table}

\subsection{Performance comparison with baselines}
\label{subsec: expt_1}

The comparison between the AFL-Net and the AVR-Net is depicted in Table~\ref{tab:main}.
It shows that the proposed AFL-Net consistently surpasses the AVR-Net on both the AVA-AVD and combined datasets. Notably, the AFL-Net trained on the AVA-AVD dataset achieves a DER of 23.65\%, significantly exceeding the baseline system with a relative DER reduction of 13.8\%, thus setting a new state-of-the-art (SOTA) performance.
Given that the MR and FAR performance remains identical between the proposed and baseline systems, the DER reduction can be attributed to a decrease in the speaker error rate. This reduction confirms that the proposed AFL-Net exhibits superior identity discrimination compared to the AVR-Net.
Moreover, by integrating large-scale datasets like VoxCeleb1 and VoxCeleb2 into the training process, both systems achieve lower SpkErr and DER performance. Consistently, the AFL-Net achieves a DER of 19.65\%, surpassing the AVR-Net with a relative DER reduction of 7.0\%.
It is also worth noting that AFL-Net experiences a modest increase in size (58.0M total, 11.6M trainable parameters) compared to AVR-Net (47.5M total, 11.4M trainable parameters), but achieves a remarkable DER reduction.
We have provided some demo audios to showcase how AFL-Net outperforms AVR-Net in challenging scenarios\footnote{\url{https://afl-net.github.io/afl-net/}}.

The comparison with DyViSE is presented in Table~\ref{tab:main2}. As no open-source codes were made available, we cite the performance reported in \cite{wuerkaixi2022dyvise} directly for comparison purposes.
It's noteworthy that even without incorporating the WavLM into our audio encoder, our model still achieves comparable results (DER: 23.65\% v.s. 23.46\% in \cite{wuerkaixi2022dyvise}), despite our audio encoder being more than 30 times smaller. After integrating the WavLM into the audio encoder, AFL-Net achieves a lower DER of 22.12\%, thereby outperforming DyViSE.

\begin{figure}[t]
\begin{minipage}[b]{1.0\linewidth}
  \centering
  \centerline{\includegraphics[width=8.5cm, height=4cm]{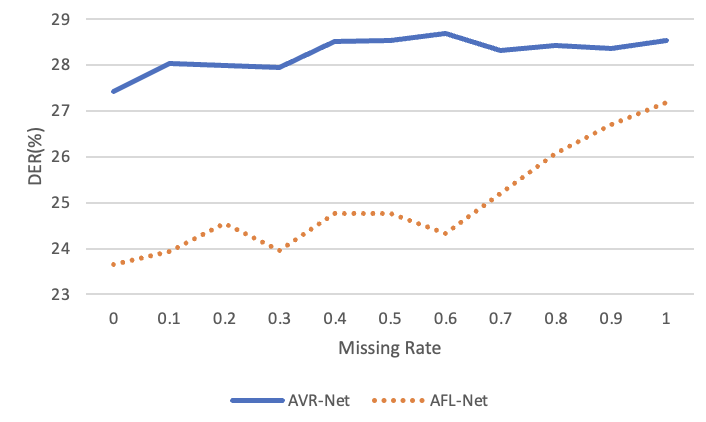}}
\end{minipage}
\caption{The DER performance comparison on AVA-AVD dataset under varying missing rates of the visual features.}
\label{fig:res}
\end{figure}

\subsection{Performance comparison under varying missing rates of the visual features}
\label{subsec: expt_2}

We carried out experiments to validate the efficacy of the AFL-Net under varying rates of visual feature absence, as shown in Fig.~\ref{fig:res}.
As the missing rate increases, both models experience a rise in DER, emphasizing the positive contribution of the visual modality to diarization.
The proposed system consistently outperforms the baseline across different missing visual feature rates, with the performance gap narrowing as the missing rate increases.
This indicates that AFL-Net surpasses AVR-Net in effectively utilizing visual information to enhance diarization performance.
This improvement can be attributed to the integration of the lip movement modality and the two-step cross-attention fusion mechanism.

\subsection{Ablation study}
\label{subsec: expt_3}

Table~\ref{tab:cross-attention-type} compares four fusion implementations for three modalities. The results indicate that all three two-step two-modality fusion strategies outperform the one-step three-modality fusion strategy. This is likely because the two-step two-modality strategies are easier for the model to learn and yield better results.
Furthermore, the best performance is achieved when initially fusing the audio and face modalities, followed by the inclusion of the lip movement modality in the second step.

An ablation study was carried out, as shown in Table~\ref{tab:ablation}. Beginning with the AFL-Net, we sequentially remove each modification in the following order: the integration of the lip movement modality, the masking strategy, and the cross-attention mechanism. It's important to note that after the removal of these three modifications, the AFL-Net reverts to the AVR-Net. From the table, we observe a consistent decline in performance following the removal of each modification, thereby validating the effectiveness and necessity of each modification.

\begin{table}[t]
\centering
\caption{Performance comparison among 4 fusion implementations on AVA-AVD dataset. In two-step two-modality fusion strategies, the modalities fused in the first step are enclosed in brackets. A+F+L denotes the one-step three-modality fusion.}
\setlength{\tabcolsep}{4.5pt}
\begin{tabular}{ccccc}
\toprule[1pt]
\multirow{1}{*}{\textbf{} }
& \multicolumn{1}{c}{\textbf{(A+F)+L}} 
& \multicolumn{1}{c}{\textbf{(A+L)+F}} 
& \multicolumn{1}{c}{\textbf{(F+L)+A}} 
& \multicolumn{1}{c}{\textbf{A+F+L} }\\ \midrule[1pt]
\textbf{SpkErr} & \textbf{21.10\%} & 22.11\% & 22.00\% & 22.59\%\\
\textbf{DER} & \textbf{23.65\%} & 24.66\% & 24.55\% & 25.14\%\\

\bottomrule[1pt]
\end{tabular}
\vspace{-1em}
\label{tab:cross-attention-type}
\end{table}

\begin{table}[h]
\centering
\caption{Ablation study on AVA-AVD and w/ Extra Data. Results are expressed in percentages (\%).}
\vspace{0.5em}
\setlength{\tabcolsep}{4.5pt}
\begin{tabular}{cllll}
\toprule[1pt]
\multirow{2}{*}{\textbf{Models}} & \multicolumn{2}{c}{\textbf{AVA-AVD}} & \multicolumn{2}{c}{\textbf{w/ Extra Data}} \\ 
& \multicolumn{1}{c}{\textbf{SpkErr} $\downarrow$} 
& \multicolumn{1}{c}{\textbf{DER} $\downarrow$} 
& \multicolumn{1}{c}{\textbf{SpkErr} $\downarrow$} 
& \multicolumn{1}{c}{\textbf{DER} $\downarrow$}\\ \midrule[1pt]

\multicolumn{1}{l}{AFL-Net}  & \textbf{21.10} & \textbf{23.65} & \textbf{17.10} & \textbf{19.65}\\
\multicolumn{1}{l}{-- lip movement} & 21.79 & 24.34 & 17.58 & 20.13\\
\multicolumn{1}{l}{-- masking}  & 22.80 & 25.35 & 18.22 & 20.77\\
\multicolumn{1}{l}{-- cross attention}  & 24.88 & 27.43 & 18.58 & 21.13 \\
\bottomrule[1pt]
\end{tabular}
\vspace{-1em}
\label{tab:ablation}
\end{table}


%% file: speaker-diarization/4-conclusion.tex
\section{Conclusion}
\label{sec:conclusion}
This work introduces the AFL-Net, enhancing AVR-Net in three key ways.
Firstly, it introduces a dynamic lip movement modality to improve identity discrimination.
Secondly, it incorporates a two-step cross-attention method for effective modality fusion.
Lastly, it proposes a masking strategy during training to encourage the system to rely more on the audio modality, which directly relates to the speaker's identity.
Experimental results consistently demonstrate that AFL-Net outperforms state-of-the-art baselines, including AVR-Net and DyViSE.